%% file: CPIS061911.tex
\documentclass[12pt]{paper}
\usepackage{amsmath}
\usepackage{amssymb}
\usepackage{amsthm}
\usepackage{pslatex}
\usepackage{ulem} 
\usepackage[pdftex]{color,graphicx}
\usepackage{cite}
\usepackage{subfigure}
\usepackage{lscape}
\usepackage{longtable}
\definecolor{MidnightBlue}{RGB}{0,64,128}
\usepackage[pdftex,colorlinks,linkcolor=MidnightBlue]{hyperref}
\usepackage{url}
\title{Robustness and Contagion in the International Financial Network}
\author{Tilman Dette$^{\dagger}$, Scott Pauls${^{@}}$,  and Daniel N. Rockmore${^{@,*}}$ \\ \small $^{\dagger}$Department of Economics,   \small Harvard University, Cambridge, MA 02138 USA\\ \small $^{@}$Department of Mathematics,    \small Dartmouth College, Hanover, NH, 03755 USA \\ \small $^{*}$The Santa Fe Institute, Santa Fe, NM 87501USA}
\begin{document}

\maketitle
\begin{abstract}
The recent financial crisis of 2008 and the 2011 indebtedness of Greece highlight the importance of understanding the structure of the global financial network. In this paper we set out to analyze and characterize this network, as captured by the IMF Coordinated Portfolio Investment Survey (CPIS), in two ways. First, through an adaptation of the ``error and attack'' methodology\cite{Albert:2000}, we show that the network is of the ``robust-yet-fragile'' type, a topology found in a wide variety of evolved networks. We compare these results against four common null-models, generated only from first-order statistics of the empirical data. In addition, we suggest a fifth, log-normal model, which generates networks that seem to match the empirical one more closely. Still, this model does not account for several higher order network statistics, which reenforces the added value of the higher-order analysis. Second, using loss-given-default dynamics\cite{Upper:2007}, we model financial interdependence and potential cascading of financial distress through the network. Preliminary simulations indicate that default by a single relatively small country like Greece can be absorbed by the network, but that default in combination with defaults of other PIGS countries (Portugal, Ireland, and Spain) could lead to a massive extinction cascade in the global economy.
\end{abstract}

\subsection*{Intro}

Globalization has created an international financial network of countries linked by trade in goods and assets. These linkages allow for more efficient resource allocation across borders, but also create potentially hazardous financial interdependence, such as the global ripple following the 2008 collapse of Lehman Brothers and the potential financial distress that may follow the potential restructuring of Greece's debt obligations. Increasingly, the tools of network science are being used as a means of articulating in a quantitative way measures of financial interdependence and stability. %  future note to self: also read up on papers by Caballero\&Simsek2011 and Acemoglu?)
%From the old CPIS paper

Network approaches have proved useful for articulating the interdependence of many kinds of complex systems, including  economic systems and the global economy (see e.g., \cite{schw2009,Garas2010}).  Of particular relevance are the studies of the World Trade Web (WTW), derived from OECD data\cite{wtw}, in which countries are linked according to the value of exports between them. %A nalysis of the WTW reveal a power law structure in the degree distribution, as well as interesting community structure \cite{s2009,serrano-2003-68,fagiolo-2007,tdr2008}. 
The WTW articulates just one aspect of the global economy and the focus herein is on the international financial asset network as derived from the CPIS, a key financial network whose assets grew by a factor of $13.5$ in real US dollars in the three decades leading up to the 2008 financial crisis\cite{bis:2008}, exceeding annual world GDP in 2006 and 2007. Our interest is in characterizing the structure of the CPIS network, especially as it relates to aspects of systemic risk related to nation default. While the related problem of bank ``contagion" in interbank networks has drawn much interest (see e.g., \cite{gai,AAE,AGKMP,BST}), the analogous problem considered at  the scale of nations in the network of inter-nation investment has attracted much less research attention, in spite of its acknowledged and reported importance for the safety and health of the global economy \cite{economist}.  Initial efforts in this direction have been made via the study of degree distributions  \cite{Song:2009, Kubelec:2009}.
%, but for various specifications of the network as derived from the CPIS, the results do not support an obvious power law (see the Supplementary Information, Figure S1) and as a consequence, there we find no a priori support for the conjectured ``robust-yet-fragile" nature. 

%The integration of new network-related  methodologies to economic situations is of great current interest \cite{F1}. The coupling of dynamics with network thinking bears some relation to the new contagion analysis that is now finding some purchase in the economics literature (see e.g., \cite{AAE,AGKMP,BST}).

%Previous network analyses of the WTW have provided a wealth of initial information regarding the interaction of the network structure and the functional architecture of the WTW.  Our goal is to extend this analysis and to craft measures of robustness or stability that take into account the intertwined structure and aspects of dynamics.  Our core methodology is that of a {\em knockout experiment}, where stability is tested via perturbation of the network by removal or degradation of some aspect of the structure, which in turn initiates some dynamic reorganization of the network.  We then observe the system as it returns to equilibrium and record the effect.

%%From the knockoutpaper
In particular, in this paper we perform two kinds of analyses. In the first we perform so-called ``knockout experiments" on a network of countries connected according to a threshold in their CPIS financial relationship. In these experiments  countries are removed one by one from the network (via some criteria explained below) and with this, all connections to and from the corresponding nodes.  Robustness is measured according to the degradation in connectivity as measured by an increase in the average shortest path length (ASPL) of the network; ASPL provides a natural proxy for financial integration.
Knockout methodology has already been applied in a variety of contexts, including the World Wide Web (WWW)~\cite{Albert:2000}, metabolic networks \cite{Jeong2000}, protein networks \cite{Jeong2001},  and in the form of {\em extinction analyses} conducted on food web models of ecosystems (see e.g., Section 4.6 of \cite{dunne2006} and the many references therein as well as  \cite{Allesina2009}). In the economic context knockouts have very recently been applied to study the WTW \cite{FPR-JEDC}.

\subsection*{The Network}

We derive the international financial network structures from the IMF's CPIS database. The IMF CPIS comprises bilateral annual data from 2001 to 2009 derived from the external portfolio of financial assets aggregated at the country level from up to $73$ reporting countries vis-\`a-vis $237$ countries\cite{CPIS:2002,cpis-data}. External assets are reported in terms of millions of current US dollars (USD) and thresholded at $500,000$ USD. If we restrict the network to reporting countries with available GDP data\cite{gdp1,gdp2}, we obtain a subset of at least $64$ countries for each year. The portfolios of these countries restricted to the same subset give a self-contained network that accounts for at least $97.4$ percent of their total external assets. For the sake of analytical consistency, we restrict our analysis to these core subsets.

\begin{figure}[htbp]
	\centering
		\includegraphics[width=\textwidth]{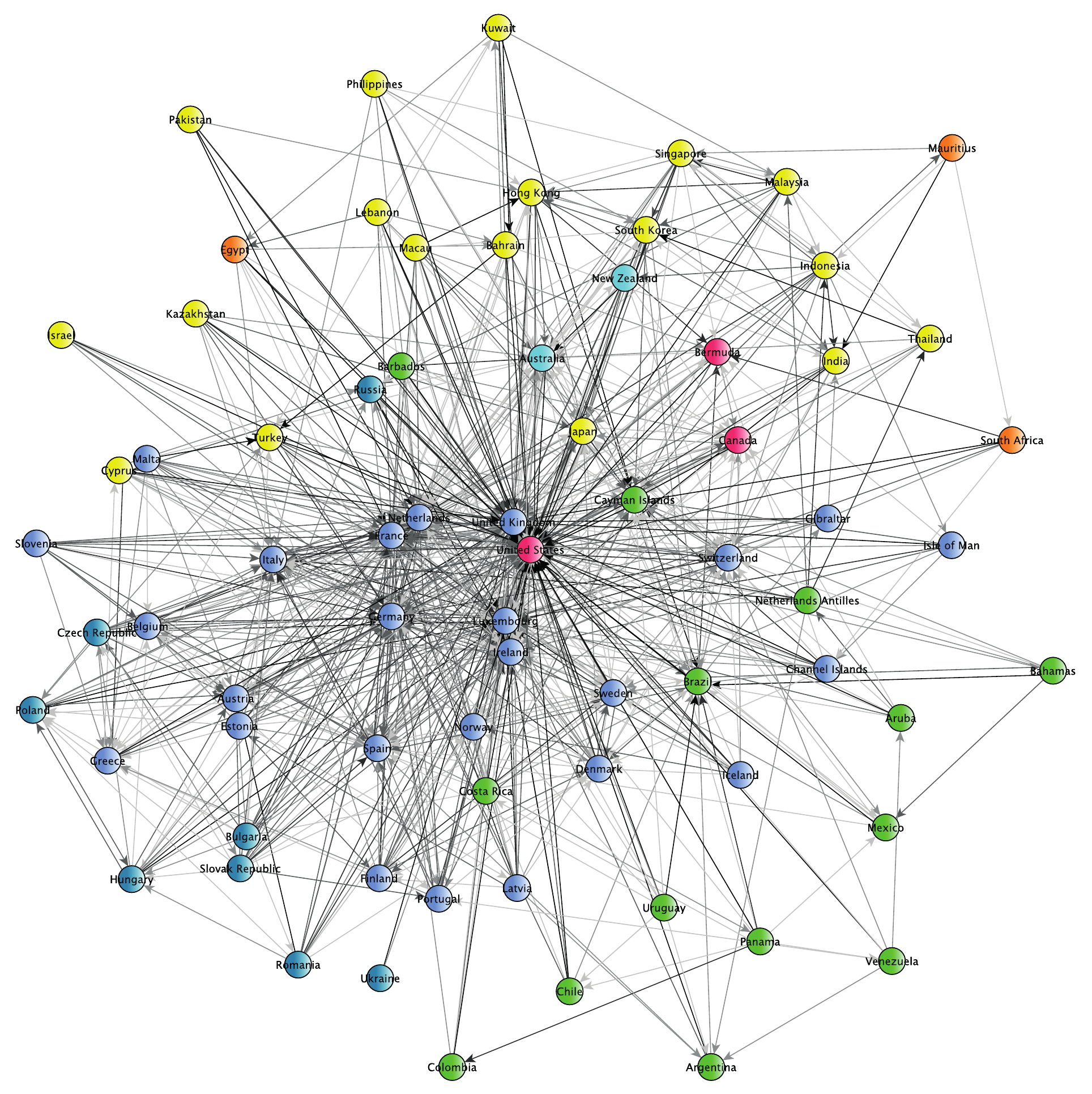}
	\caption{\textbf{The international financial network in 2009} as derived from the CPIS data. The graph shows a directed edge from country $i$ to country $j$, if $i$'s asset holdings w.r.t\ $j$ exceeds its average exposure to other countries. Successively darkened edges indicate a corresponding exposure that is at least $2\times$, $4\times$, $8\times$ or above $16\times$ the average exposure. Nodes are color coded by different geographical regions as defined by the UN. } %***CAN YOU GIVE THE COLOR CODE HERE?**
	\label{fig:figs_A_in_2009}
\end{figure}

We encode a principal binary network structure for these core subsets in two adjacency matrices $A,$ and $B$, by applying two different thresholding rules. For a given year, let $s_{ij}$ denote the amount of country $i$'s assets issued by residents of country $j$ and let $n$ be the number of countries in the core subset under investigation. Then, matrix $A$ encodes an edge ($a_{ij}=1$) if $s_{ij}>\sum_{k\not=i}\frac{s_{ik}}{n-1}$, that is, if country $i$'s portfolio contains above average exposure to country $j$. See Figure \ref{fig:figs_A_in_2009} for a visualization of the representation of $A$ for the 2009 data. Matrix $B$, on the other hand, encodes an edge ($b_{ij}=1)$ if $s_{ij}/GDP_{i}$ exceeds a threshold $t=.0417$, representing the average $\% 4.17$ percent GDP-normalized investment in these networks. We choose these two simplified representations of the network,  but acknowledge the hidden complexity that is missed in the aggregation of assets in the CPIS data and simplified international financial links. Future work should work to articulate the finer detail; we see this paper as a first and necessary step toward a more sophisticated analysis.

Given the binary directed networks, we generate comparison networks according to four common null-models. For the simplest, Erd\'{o}s-Renyi (ER) model, $G^{1}\left(n,p\right)$ has $n$ nodes and an edge between any two nodes exist with probability $p=\bar{d}/\left(n-1\right)$, where $\bar{d}$ is the average out-degree of the empirical network. For the second and third models, $G^{2}\left(n,\left[p\right]_{i}\right)$ and $G^{3}(n,[p]_{j})$ have an edge between countries $i$ and $j$ with probabilities $p_{i}=d_{i}^{out}/\left(n-1\right)$ and $p_{j}=d_{j}^{in}/\left(n-1\right)$, where $d_{i}^{out}$ and $d_{j}^{in}$ are country $i$ and $j$'s empirical out- and in-degree. Finally, the fourth model assigns equal probability to all graphs $G^{4}\left(n,\left[in\right]_{i},\left[out\right]_{i}\right)$ that preserve the in- and out-degree of $A$. To generate such graphs, we use the rewiring approach (see e.g., \cite{Maslov03052002}). Thus, these four null models each generate networks based only on an increasing number of first order characteristics of the empirical binary network.

\begin{figure}[htb]
	\centering
		\subfigure[Empirical Data]{
			\includegraphics[width=.47\textwidth]{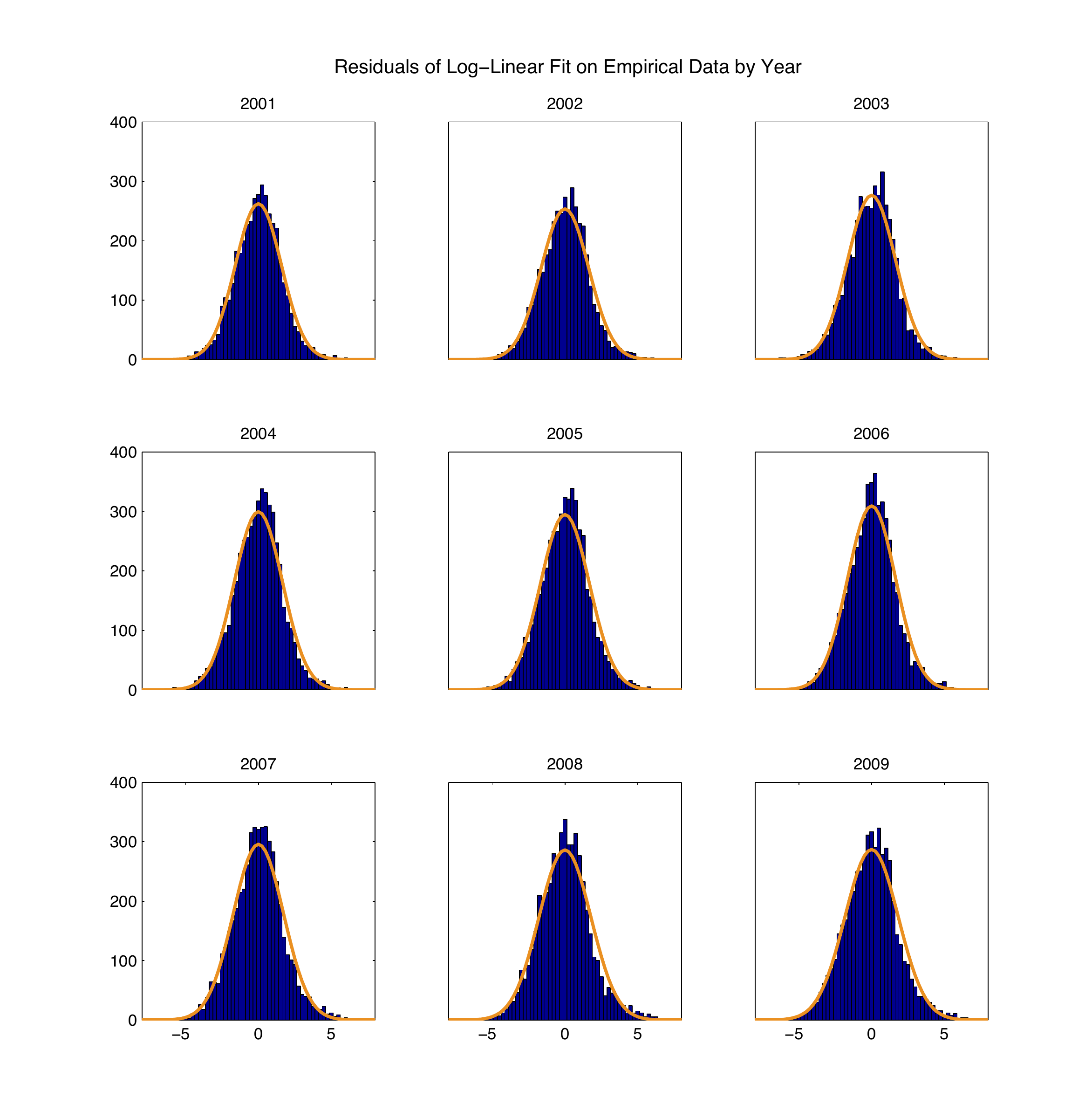}
			\label{fig:figs_log_norm_dist_empirical}
		}
		\subfigure[Generated Data]{
			\includegraphics[width=.47\textwidth]{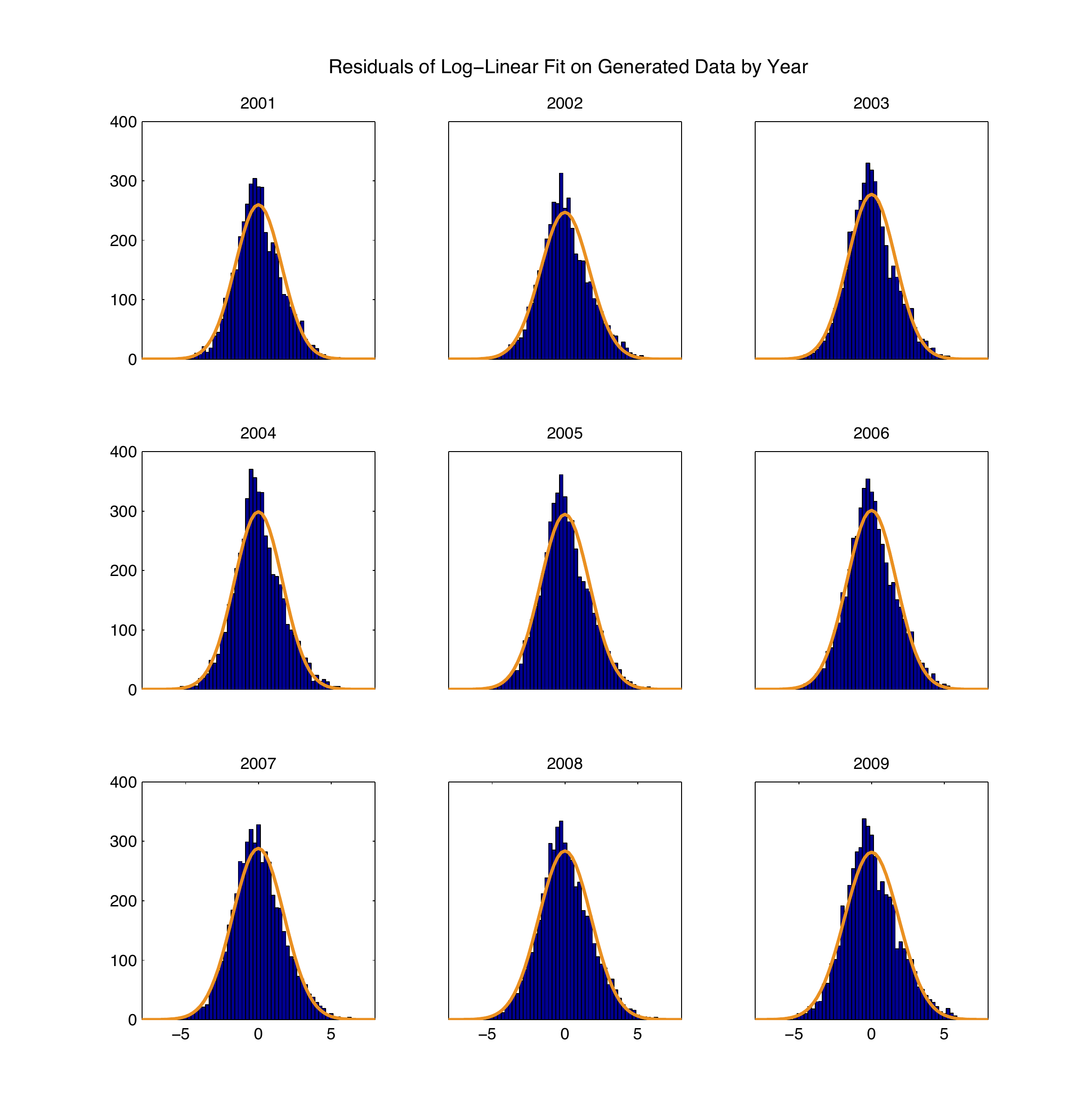}
			\label{fig:figs_log_norm_dist_generated}
		}
	\caption{\textbf{Distribution of residuals for 2001-2009} with overlaid best fit normal distribution after fitting the log-linear model of (\ref{eqn:lnfit}) to the empirical data and corresponding generated data. }
\end{figure}

We also make use of a fifth null comparison model that takes into account the weight distribution of the original asset network. We find a reasonable fit to the asset distributions by using a log-normal model with country dummies
\begin{equation}
	\ln\left(s_{ij}+1\right)=\alpha_{i}+\beta_{j}+\varepsilon_{ij},\label{eqn:lnfit}
\end{equation}
where $\alpha_{i}$ and $\beta_{j}$ are constants for each country as q holder and issuer of assets, and $\varepsilon\sim\mathcal{N}\left(0,\sigma\right)$. The plus one term on the LHS is a work-around, given that the CPIS data is left censored. We use maximum-likelihood estimation to predict the $\alpha$'s, $\beta$'s and $\sigma$.%
\footnote{Noting that the left-censoring and rounding implicit in the CPIS data downward biases the $\sigma$ estimate, we scale this estimate by 1.183, as suggested by fitting the model to generated but left-censored and rounded data.}
Figure \ref{fig:figs_log_norm_dist_empirical} shows the distribution of predicted $\varepsilon_{ij}$ after fitting this model to all nine years of available data. While a Jarque--Bera test rejects that these residuals are normally distributed for several years, this test similarly results in a type I error when applied to data generated by the log-normal model but was left-censored and rounded. Comparing the predicted residual distributions of the empirical and generated data visually provides further evidence that this simple model fits the empirical data surprisingly well. 
% (THE SENSE OF THIS PARAGRAPH IS CONFUSING -- IS THIS A REASONABLE MODEL OR NOT?) -- It is, as the prediceted error distributions of both emprical and generated data seem surprisingly similar.

\subsection*{Error and Attack}

We test the robustness of the CPIS network as captured by matrices $A$ or $B$ via the effect of ``error and attack'' simulations \cite{Albert:2000} on the average shortest path length (ASPL) in the network. Here the network is subjected to the iterated removal of either a random node (via ``error") or the most ``important'' (via ``attack"), as measured by the sum of a node's in- and out-degree. While other measures of importance can also be used, our measure follows intuitively as this `attack' takes out the greatest number of direct paths, thus attacking the  tracked ASPL measure.

\begin{figure}[htbp]
	\centering
		\includegraphics[width=\textwidth]{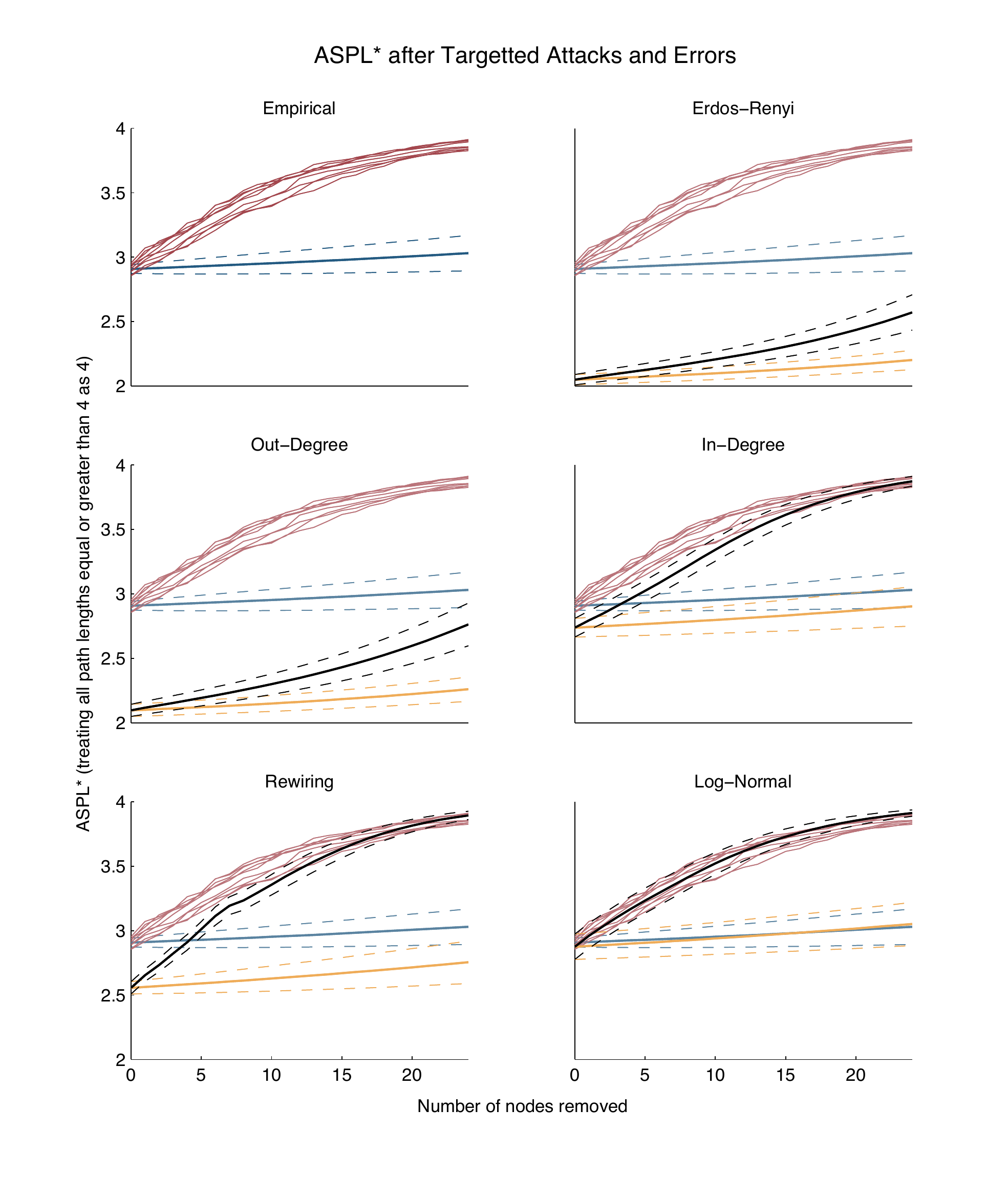}
	\caption{\textbf{Evolution of ASPL - A} -- Plots summarize the error and attack analysis for the empirical network of type A and five null comparison models for 9 separate years of data. The first plot shows the evolution of the modified ASPL of all 9 empirical networks under targeted attack (red) and the general evolution under random errors (blue) showing the mean and standard deviations of $9\times2000$ independent simulations. Subsequent plots superimpose the general evolution of $9\times2000$ generated networks under error (yellow) and attack (black) by null model.}
	\label{fig:figs_error_attack_A_typo}
\end{figure}

\begin{figure}[htbp]
	\centering
		\includegraphics[width=\textwidth]{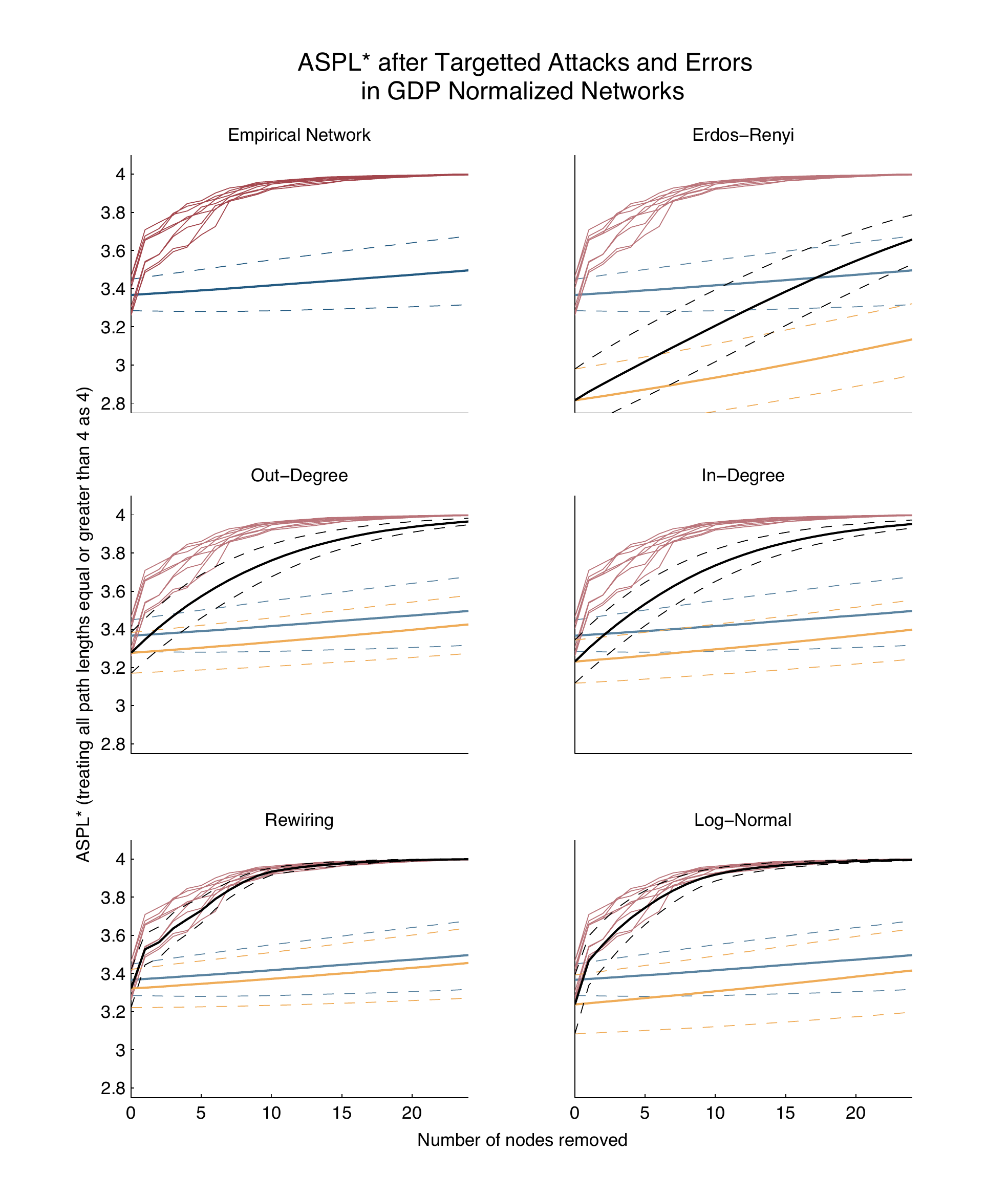}
	\caption{\textbf{Evolution of ASPL - B} -- See description of figure \ref{fig:figs_error_attack_A_typo}. }
	\label{fig:figs_error_attack_B_typo}
\end{figure}

The shortest path length (SPL) between any two nodes is a useful proxy for financial integration; a low SPL indicates a high degree of direct investment (a shortest path length of one) and common investment through cross-border positions of intermediate countries (a shortest path length of two). However, as the length of the SPL grows it becomes a less useful indicator. Also, not all pairs of nodes have paths between them. Thus we choose a modified measure of the ASPL, by treating all SPLs greater than three as having a length of four.%
\footnote{We also tracked the evolution of simpler measures, such as the fraction of SPLs equal or less than two. These measured provided qualitatively very similar results.%
} Figures \ref{fig:figs_error_attack_A_typo} and \ref{fig:figs_error_attack_B_typo} show the general evolution of the modified ASPL under repeated error or attack for the empirical as well as the null model networks, given the two specifications.

% \textbf{Here is my understanding of the figures in Figure 3: The figure in the upper left is embedded in all the figures. In red are the attack studies for each of the nine years of data. In blue is the average random attack with error bars given by the dashed lines. In each of the other figures we overlay the attack and the error studies performed on the relevant null models - again with error bars. In these the black is the "attack" and the orange is the "error" - is that right? As per my accompanying note, the take away message needs to be a little better explained.}

The simulations reflect the existence of key financial centers. While random removal of nodes does not noticeably affect ASPL, targeted removal causes an immediate and rapid increase for both types of thresholding. The simulations on networks of type $A$, point to the importance of financial hubs. The empirical network starts out significantly less `integrated' than the first four null models. Further, the attack method affects ASPL significantly less in the first two null models which disregard the in-degrees of countries in the empirical model. As simulations on the third and fourth null models show, matching the empirical in-degree sequence is still insufficient to explain the networks lack of greater integration as proxied by ASPL; though, the log-normal model appears to match the empirical results very well.

For simulations on the GDP thresholded $B$ networks, the rewiring null model appears to match the empirical model best. Again, the empirical model starts less integrated than the first three null models and is particularly susceptible to the first attack, the removal of the worlds largest economy, the US. The rewiring and the log-normal model both seem to match these results closely, with the former doing slightly better. Thus, it appears the observed error-attack effect on the networks' ASPL may already be encoded in the first order statistics used for the null model.

We formally test whether the ASPL, as well as several other higher order statistics of the empirical networks, are outliers within the null model families. For each empirical network we generated 10000 networks of each null-model and constructed 95 percent confidence intervals for the ASPL and 5 other network statistics. Table \ref{tab:net_measures}) summarizes how often the empirical networks produced measures below or above these confidence intervals. The results support several of our above conjectures; e.g we find that the ASPL of the $B$ networks is indeed best matched rewiring null model. Still, none of the null models can properly account for all the listed higher order statistics. Notably, the last measure, the probability that country $i$ has a path to $k$ conditional on $i$ having path to $j$ and $j$ a path to $k$ is above the confidence intervals of all null models for all years and specifications. Hence, the null models' first-order statistics appear unable to account for several relevant characteristics of the empirical network structure.

\begin{table}
	\begin{center}
		\small
		\begin{tabular}{l|rr|rr|rr|rr|rr}
		\hline\hline
		&\multicolumn{2}{c|}{ER}&\multicolumn{2}{c|}{Out-deg.}&\multicolumn{2}{c|}{In-deg.}&\multicolumn{2}{c|}{Rewiring}&\multicolumn{2}{c}{Log-Normal}\\
		Network Measure     & A & B & A & B & A & B & A & B & A & B \\
		\hline
		Fraction of SP $\le 2$      & -1 & -1    &	-1 & 0       &	-.79 & -.33  &	-1 & -.56    &	0    & -.11  \\
		Fraction of SP $\le 3$      & -1 & -1    &	-1 & -.33    &	-1   & -.78  &	-1 & -.56    &	-.11 & -.67  \\
		Modified ASPL               & 1  & 1     &	1  & .33     &	.89  & .78   &	1  & .67     &	0    & .44   \\
		Assortativity               & -1 & -1    &	-1 & -.89    &	1    & -.89  &	1  & .78     &	.11  & 0     \\
		Avg.\ Clustering Coefficent & 1  & 1     &	1  & 1       &	1    & 1     &	1  & -.22    &	1    & 0     \\
		Pr$(i\rightarrow k\mid i\rightarrow j \land j\rightarrow k)$ 
		             & 1  & 1     &	1  & 1       &	1    & 1     &	1  & 1       &	1    & 1     \\
		\hline\hline
		\end{tabular}
	\end{center}
	\caption{\textbf{95\% Confidence Interval for Network Measures} -- The table shows the fraction of years in which the empirical network measure was below the 95 precent confidence interval (negative) or above (positive) for each null model and network type. Modified ASPL provides the average shortest (directed) path length, capping maximum path length at four. Assortativity and average clustering coefficient follow the Matlab algorithms of the Brain Connectivity Toolbox.\cite{bct}}
	\label{tab:net_measures}
\end{table}

\subsection*{Preliminary LGD Simulations}

The error and attack methodology does not account for any potential dynamics -- that is, the simulation proceeds simply by successive deletion of nodes without any accounting for the potential contagion of financial distress. In our second set of simulations we attempt to model such potential economic dynamics, that is, financial interdependence. We adopt so-called loss-given-default (LGD) dynamics, originally developed for modeling cascades of bank failures \cite{Furfine:2003,Upper:2007}, to provide a method of propagating financial shock. The simulation starts with an initial country or set of countries defaulting on their financial liabilities. Any other country whose financial positions in the defaulted country/countries exceeds certain thresholding conditions will also default, which in turn may result in further defaults. The simulation terminates when no further countries are defaulting. Analogous extinction simulations are common as a means for understanding the robustness of food webs \cite{Allesina:2009}, ecological networks whose nodes represent species in an ecosystem wherein species $i$ is linked to species $j$ if $j$ preys on $i$ (so that $i$ provides resources to $j$) \cite{Allesina:2009,dunne2006,d2004}. More recently, these kinds of extinction studies have been analogized to the WTW \cite{FPR-JEDC}. 

We acknowledge that such simulations may be overly simplistic, given the degree of aggregation of the CPIS data and the great heterogeneity of international banking activities. Nonetheless, they provide an interesting start for modeling potentially severe global cascades of financial distress. As a result, the following analysis is still preliminary.

For the LGD simulations, we specify two separate thresholding conditions. A country will default if its total investment in the defaulted countries both exceeds some fraction ($d_{1}$) of its total external investment and some fraction ($d_{2}$) of its GDP; that is, country $i$ will default if $\sum_{j\in D}a_{ij}>d_{1}\sum_{j=1}^{N}a_{ij}$ and $\quad\sum_{j\in D}a_{ij}>d_{2}GDP_{i}$, with D being the set of defaulted countries. As a result, threshold $d_{1}$ incorporates a general ability of countries to absorb sufficiently small losses relative to its portfolio and threshold $d_{2}$ assumes a country-specific ability to absorb losses of assets proportional a country's economic output. Note, we may incorporate a fact, that a default us unlikely to yield a complete loss of financial assets of the defaulting country (but a certain haircut) by scaling $d_{1}$ and $d_{2}$ appropriately.

\begin{figure} 
   \centering
   \begin{tabular}{c|c|c|c}
      \hline\hline
      \includegraphics[height=.3\textwidth]{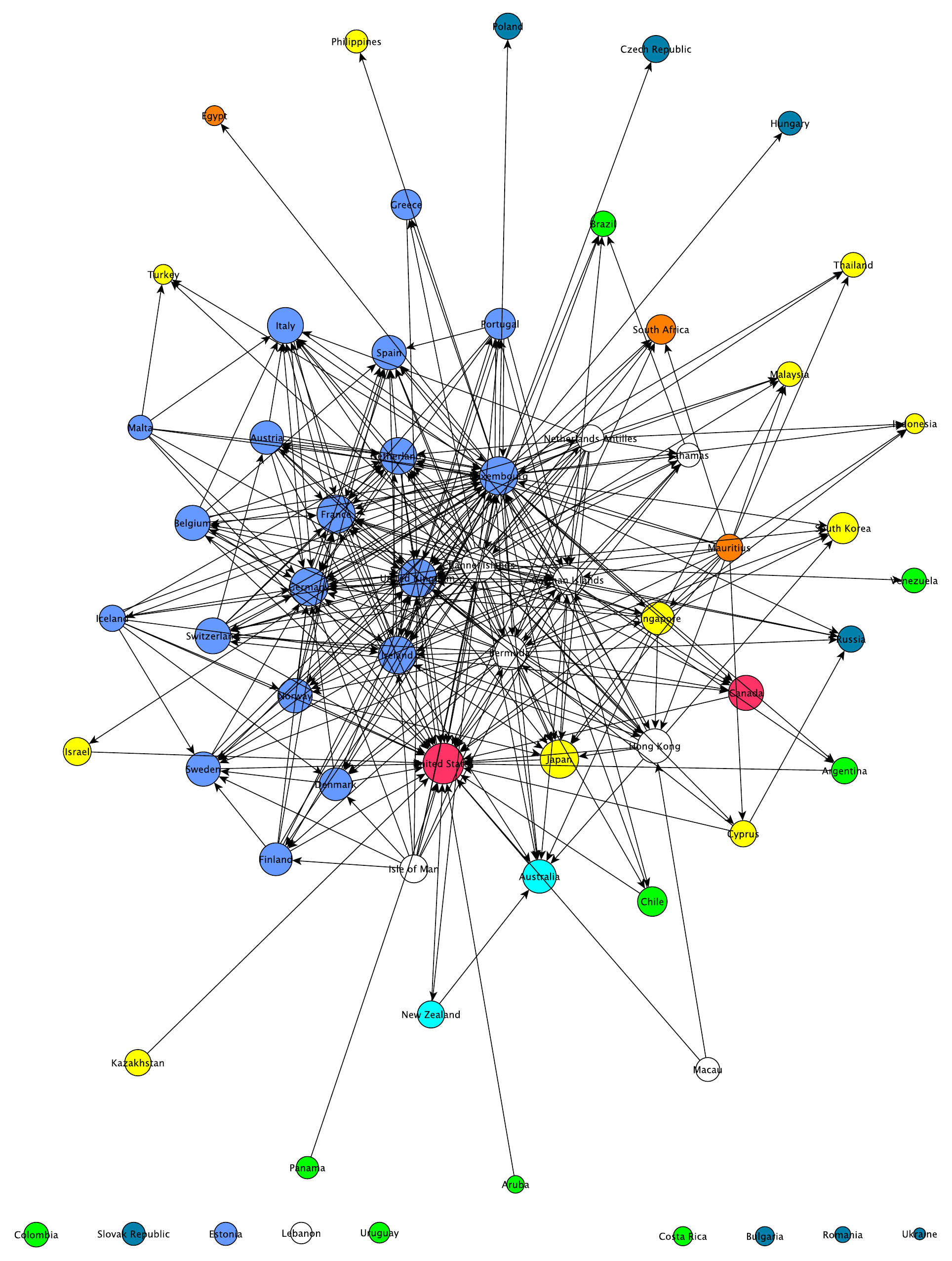} &
      \includegraphics[height=.3\textwidth]{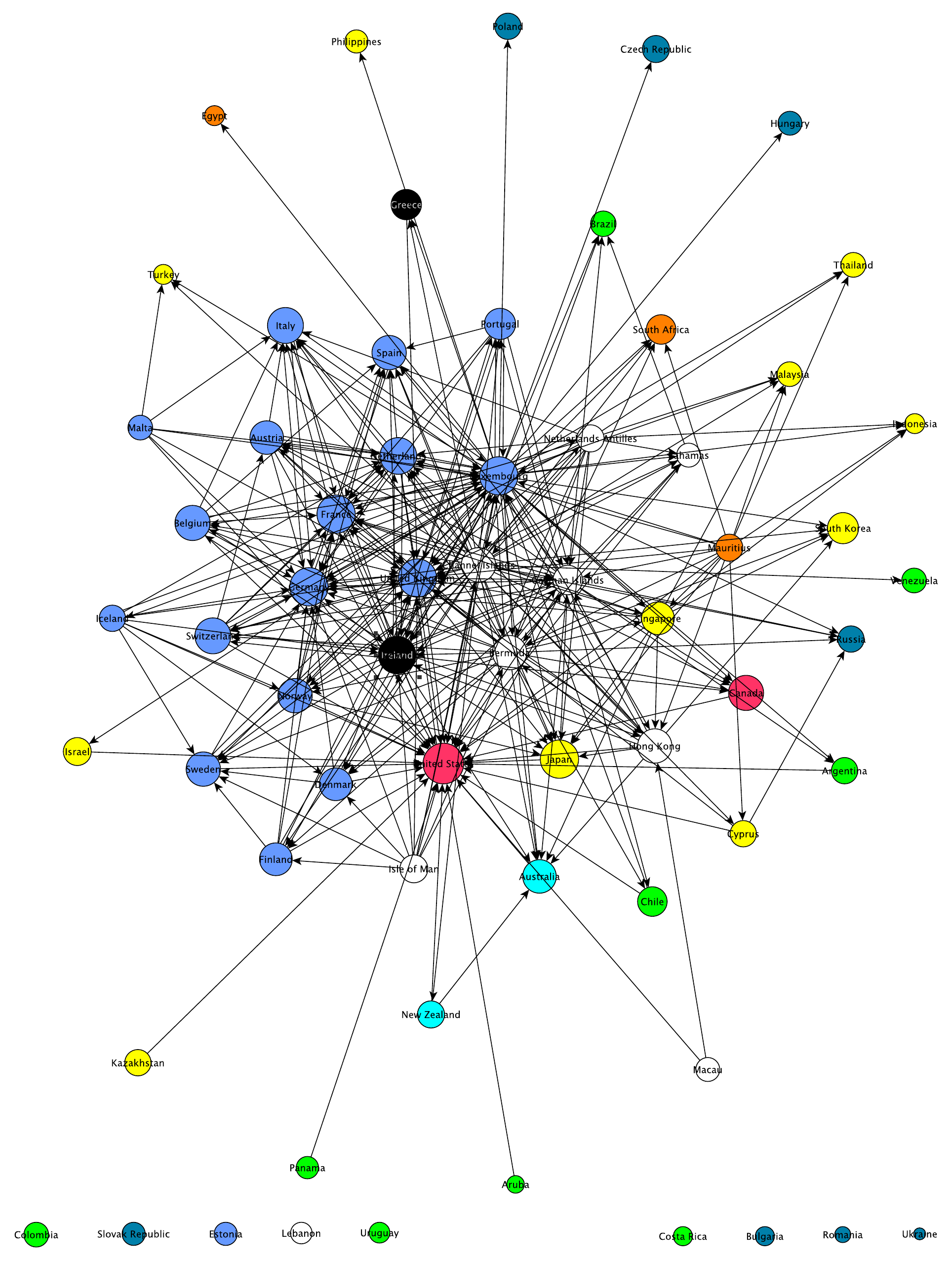} &
      \includegraphics[height=.3\textwidth]{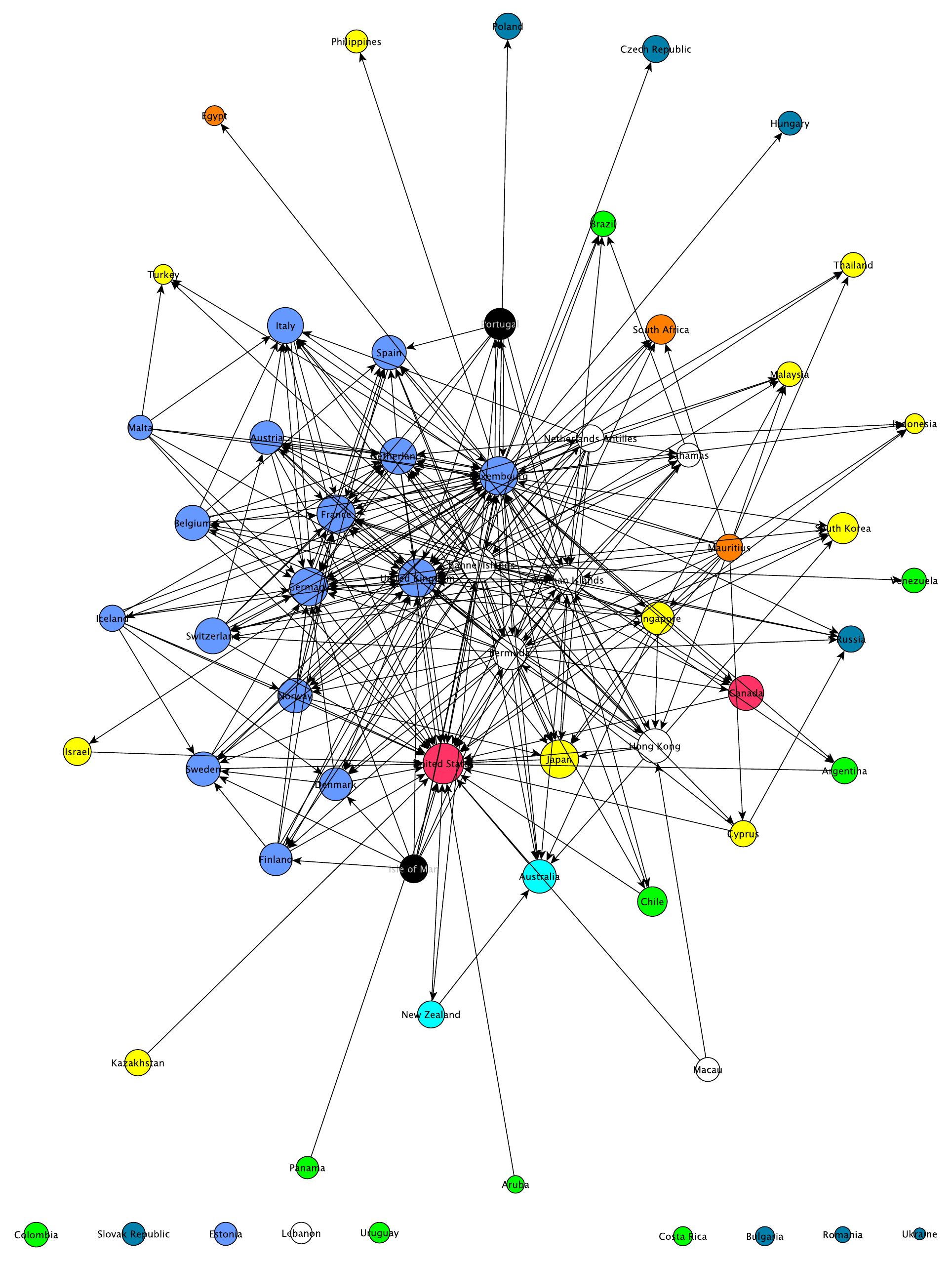} &
      \includegraphics[height=.3\textwidth]{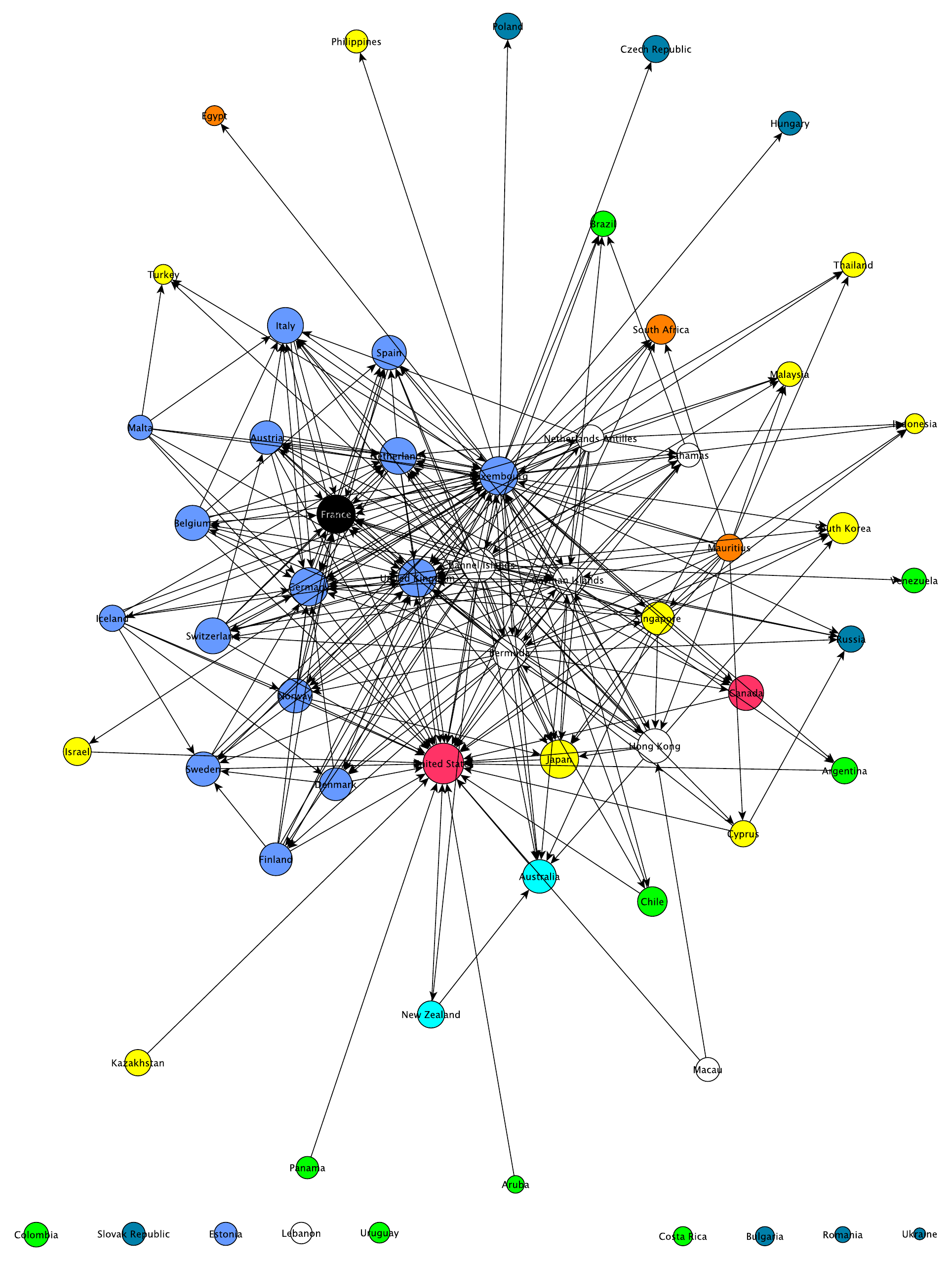}\\
      \hline
      \includegraphics[height=.3\textwidth]{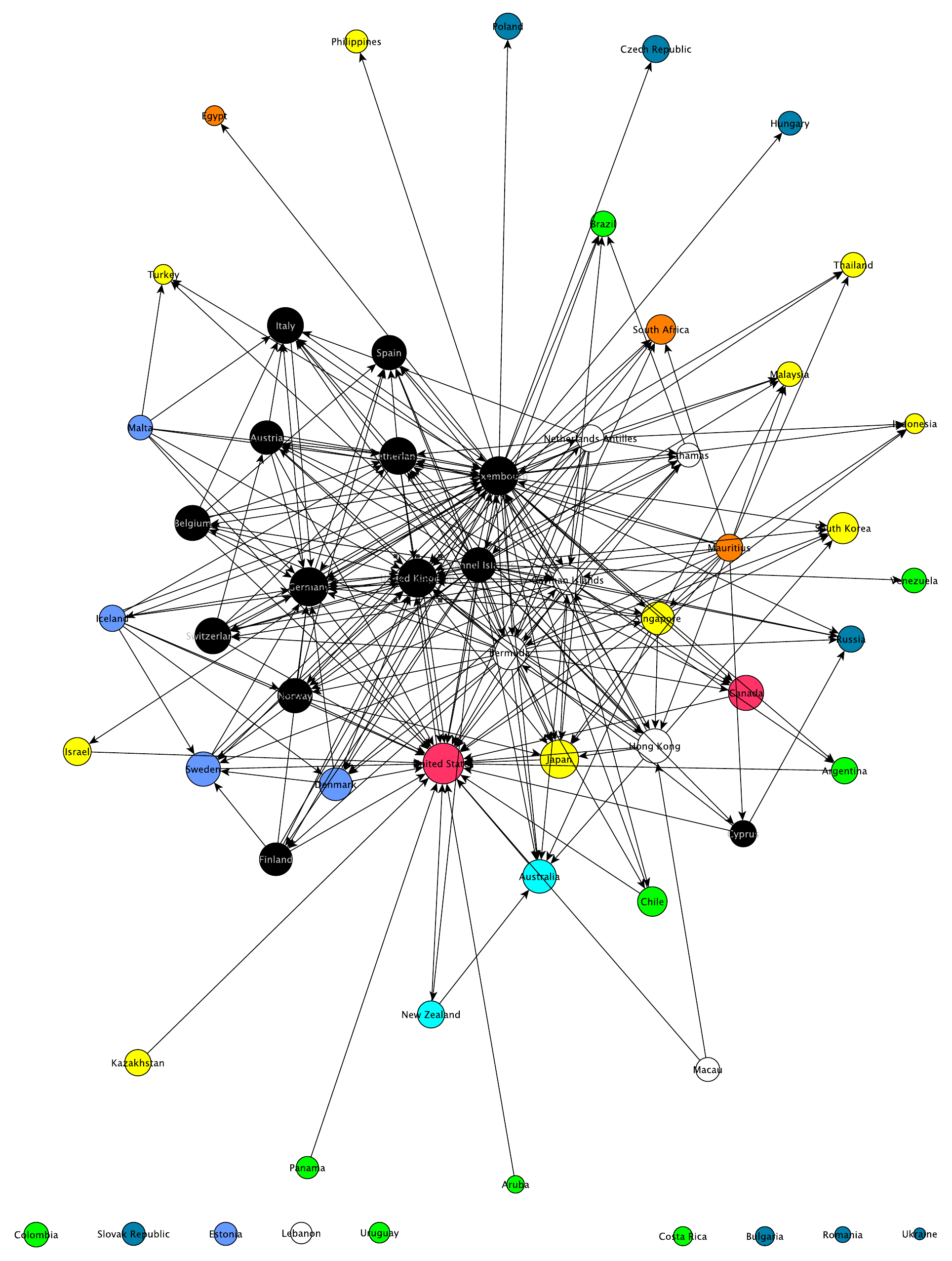} &
      \includegraphics[height=.3\textwidth]{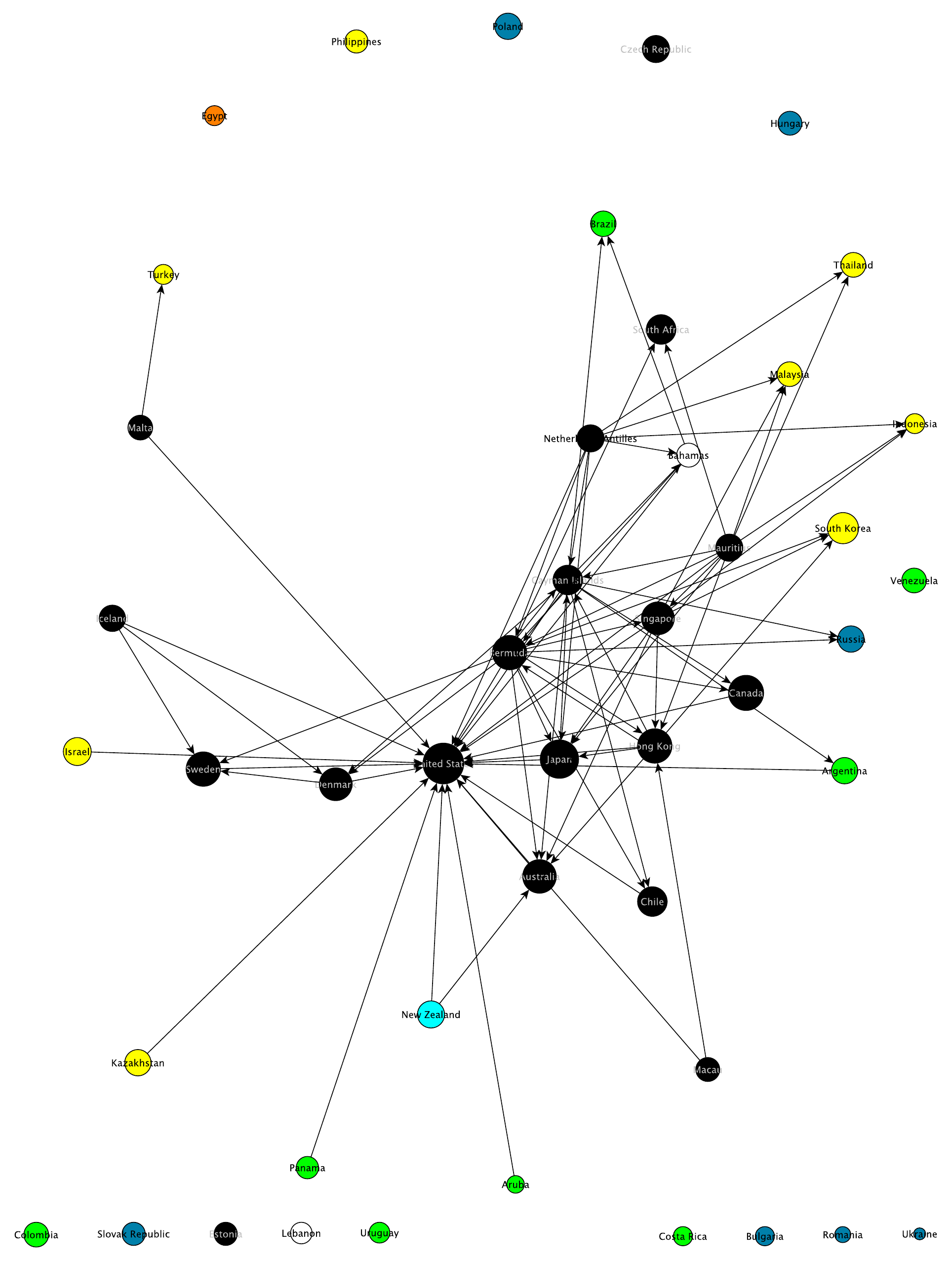} &
      \includegraphics[height=.3\textwidth]{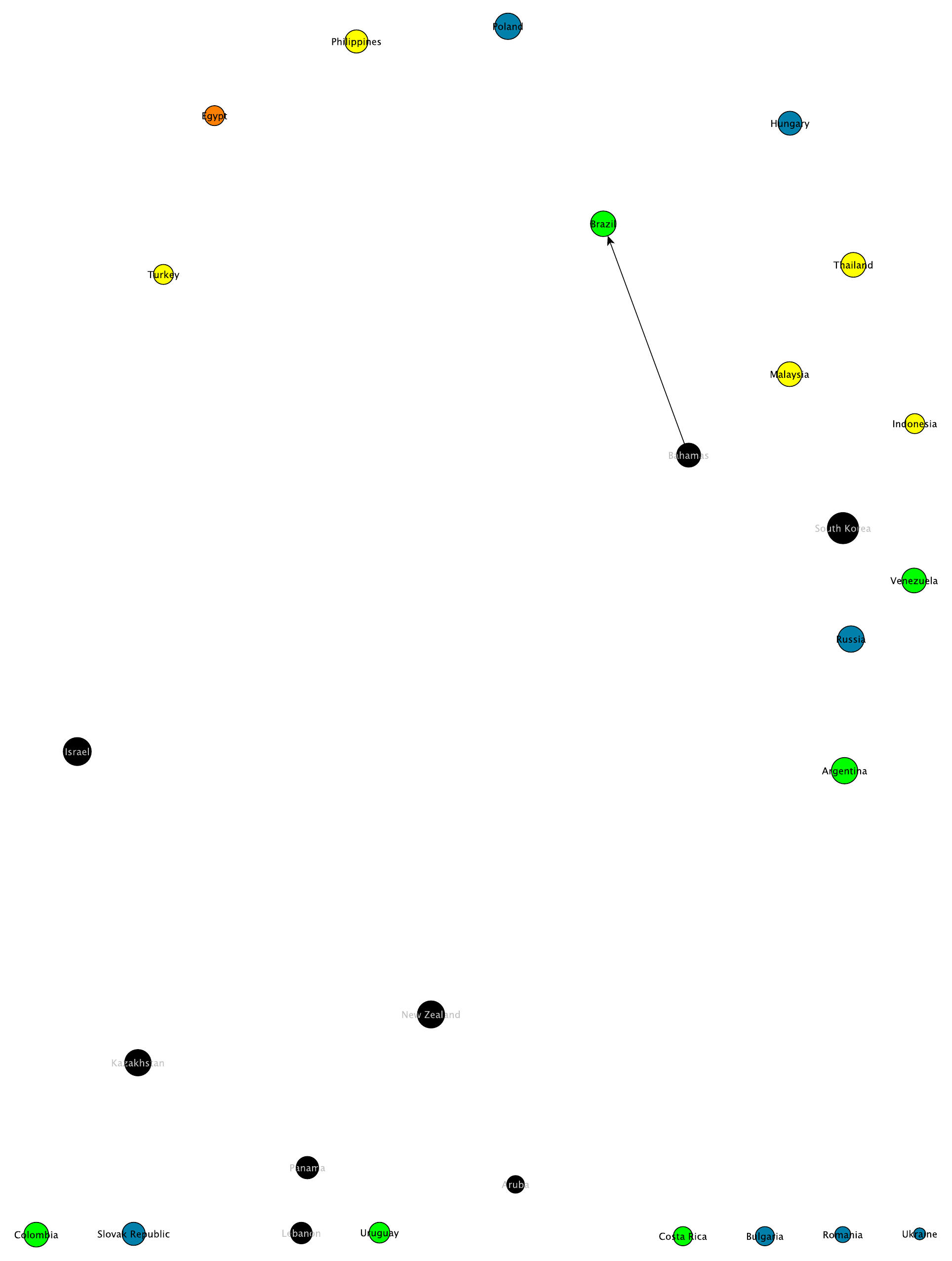} &
      \includegraphics[height=.3\textwidth]{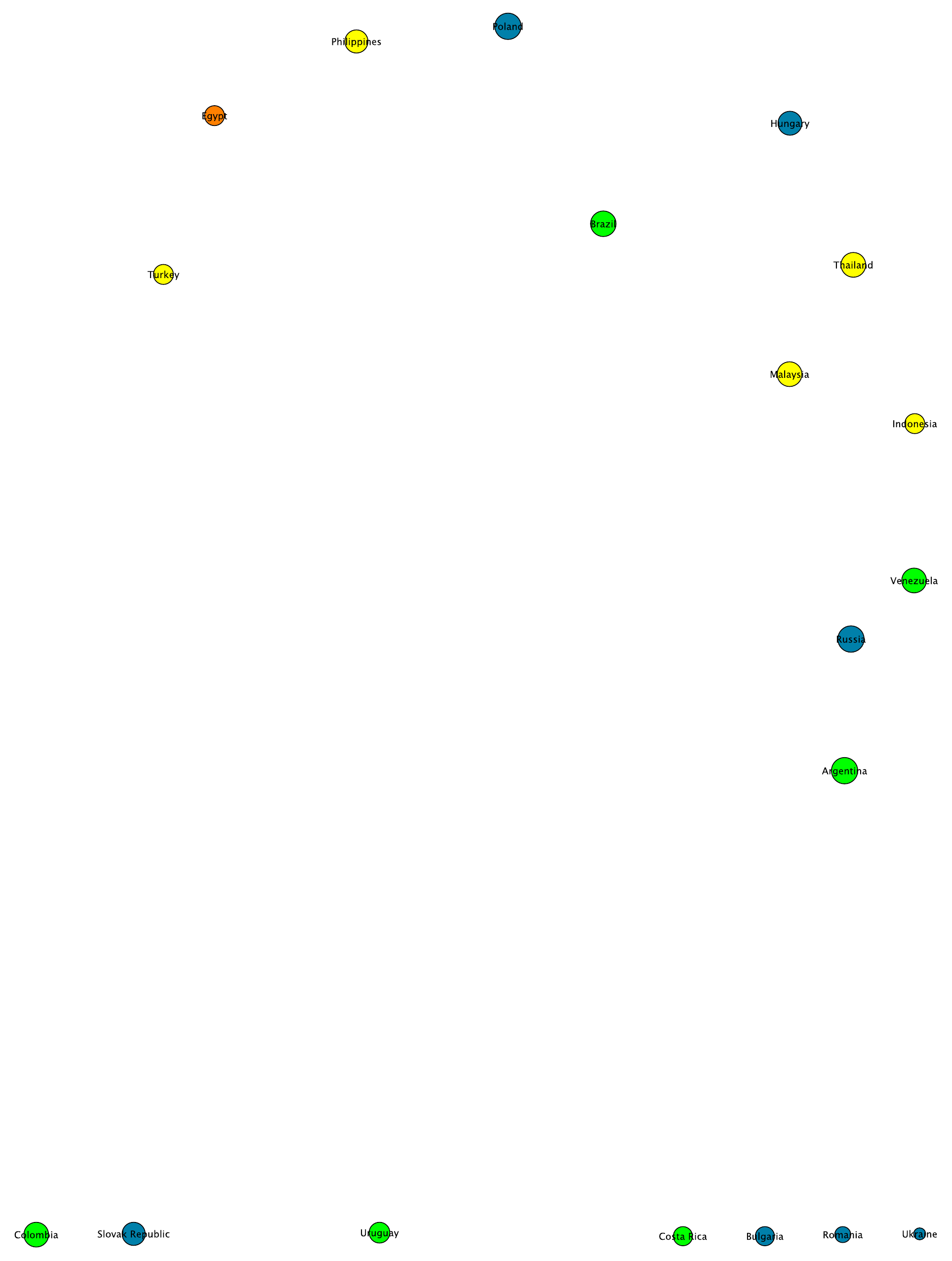} \\
      \hline\hline
   \end{tabular}
   \caption{\textbf{Exogenous Default of Greece and Ireland} -- The eight graphs show the iterative default that follows an initial default of Greece and Ireland within the 2007 financial asset network, assuming threshold condition $d_1=d_2=0.1$. At each step defaulting countries are colored black before being removed in the following graph. For graphical clarity only edges of the thresholded network are displayed, that is, an edge represents asset exposure exceeding $5.81\%$ (the average exposure) of the lending country's GDP.}
   \label{fig:figs_GRE_IRE_default}
\end{figure}

As an example, if we choose $d_{1}=d_{2}=0.1$ and initialize by defaulting any one of the PIGS countries in the 2007 network, there is at most a single subsequent default. However if we initialize by defaulting both Greece and Ireland (while using the same threshold values) then we see (Figure \ref{fig:figs_GRE_IRE_default}) six subsequent rounds of defaults in a sequence that reflects financial interdependence spreading initially across Western Europe, which subsequently affects the US and the rest of the world. In the end, only a subset of mostly emerging market economies survives the default cascade, as their total amounts of international financial assets are small relative to their respective GDP.

% Figure 2: (Greece and Ireland cascade figure)

\begin{figure}[htbp]
\centering \includegraphics[height=3.5in]{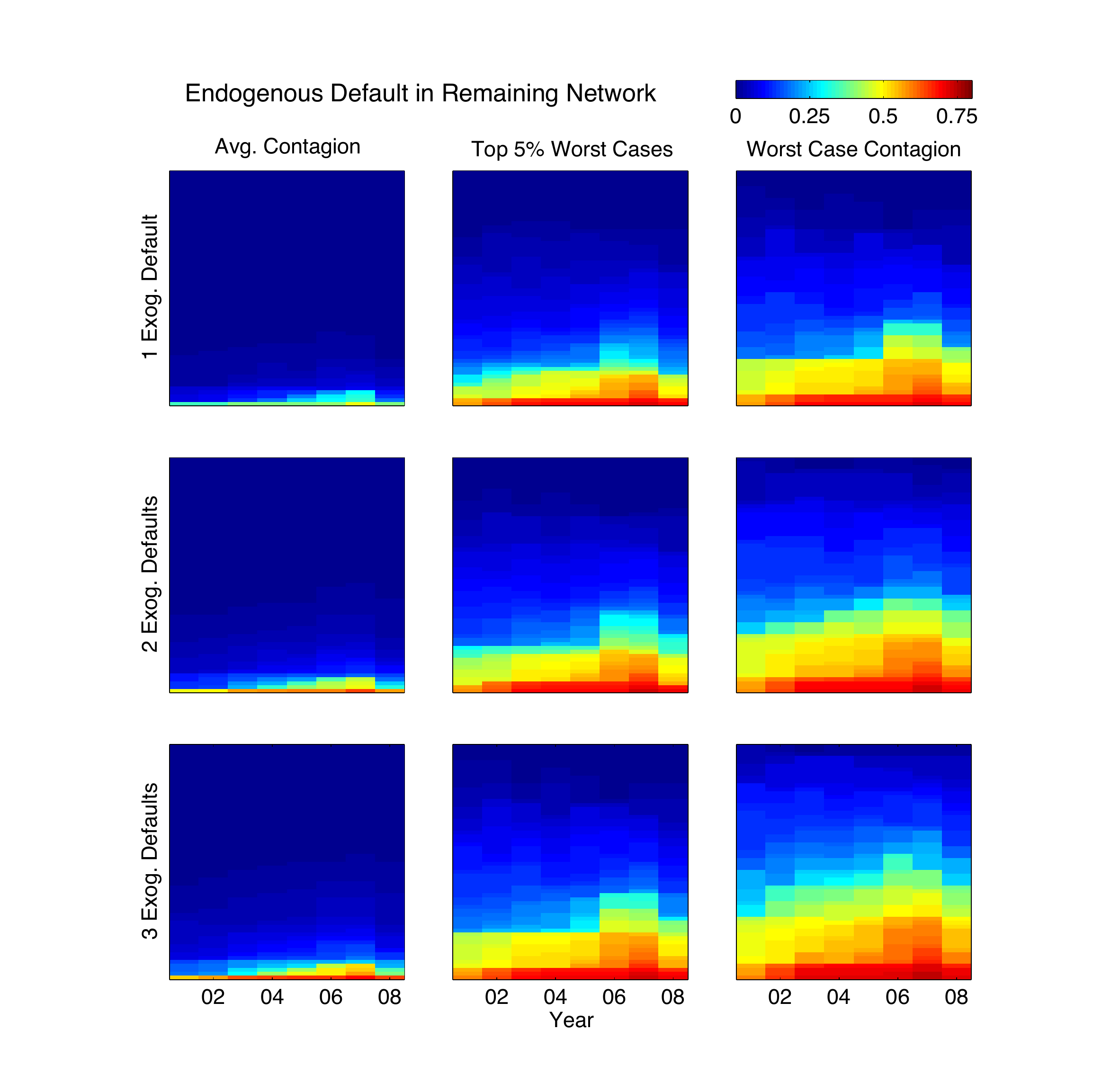}
\caption{\textbf{Summary of financial interdependence for various LGD models and given years.} For any possible combination of one, two or three exogenous defaults. In each year the plots summarize the resulting endogenous default vertically sorted by their severity (measured according to the fraction of countries that succumb to default), so that the values for $d_{i}$ are not constant along rows.}
\label{fig:4_Heat_Map_of_average_contagion} 
\end{figure}

%(Finding 2)
A broad set of LGD simulations under a range of values for $d_{1}$ and $d_{2}$ shows a notable increase in financial interdependence (as external positions increased) from 2001 to 2007. We consider all possible model specifications with $d_{1},d_{2}\in\{0,0.1,0.25,0.5,0.75\}$, excluding the trivial case $d_{1}=d_{2}=0$. For each threshold specification and year, all possible combinations of the initial default of $1,2,$ and $3$ countries were simulated, to produce the measures of the mean, mean of the worst 5\%, and worst-case impact on $A$. Impact is measured in terms of the fraction of countries that eventually default under LGD. The analysis shows a general increase in severity of worst-case contagion from 2001 to 2007; most notably, the number of $d_{1},d_{2}$ specifications that produce a default of more than $55\%$ of the network in the worst case doubles in 2006 and 2007 (see Figure \ref{fig:4_Heat_Map_of_average_contagion}). The LGD simulations show a decrease in financial interdependence at the end of 2008. The result may appear counterintuitive. However, the 2008 crises yielded a substantial amount of asset losses, and an overall reduction in country-to-country foreign asset positions. Thus, financial interdependence appears lower at the end of 2008, as the LGD model does not account for already incurred losses. Incorporating such dynamic aspects may be one area for improvement of our models. Similarly, simulations could benefit from improved foundations for thresholding conditions. Average contagion remains very limited over all years, pointing once more to the general robustness of the network.

%Figure 3: (heat maps for summery of findings?)

The simulations also identify the countries responsible for worst-case impact. Table \ref{tab:Top_countries} lists the ten most influential countries and ten most influential combinations of two or three countries as measured by their worst case default scenarios. The US is the single most contagious country followed by financial centers like the UK, the Cayman Islands and Luxembourg (which is tied with Germany for fourth place). More interesting is the frequent appearance of Brazil, as a partner (with the United States) in the second most influential pair of countries and then its prevalence among the most influential triples. Similarly, middle income countries like Turkey and Indonesia also appear in the top 10 list of influential pairs or triples. Note, that western countries tend to be strongly linked with the US, while financial exposure of emerging market economies relative to their GDP tend to substantially lower. Subsequently, in model specification where a default of the US is sufficient to bring down the western world, countries like Brazil gain in importance; they can still spur further default within the small remaining network of emerging market economies, noted above. 
%  \emph{> Simulations highlight potential `secondary' level stability of network? (I'm not sure how to formulate this, yet. Notably, emerging market economies like Brazil seemed to have weathered the 2008 crisis quite well.)}  

% (Table on who caused worst case)
\begin{table}
   \begin{center}
      \resizebox{\textwidth}{!}{
      \begin{tabular}{c}
         \input{Top_countries.tex}
      \end{tabular}
      }
   \end{center}
   \caption{\textbf{Top 10 Most Influential Countries} -- The top 10 countries and combination of countries are ranked by the number of times they resulted in the greatest LGD in the 576 simulations.}
   \label{tab:Top_countries}
\end{table}

\begin{figure}[htbp]
    \centering
       \includegraphics[height=3.5in]{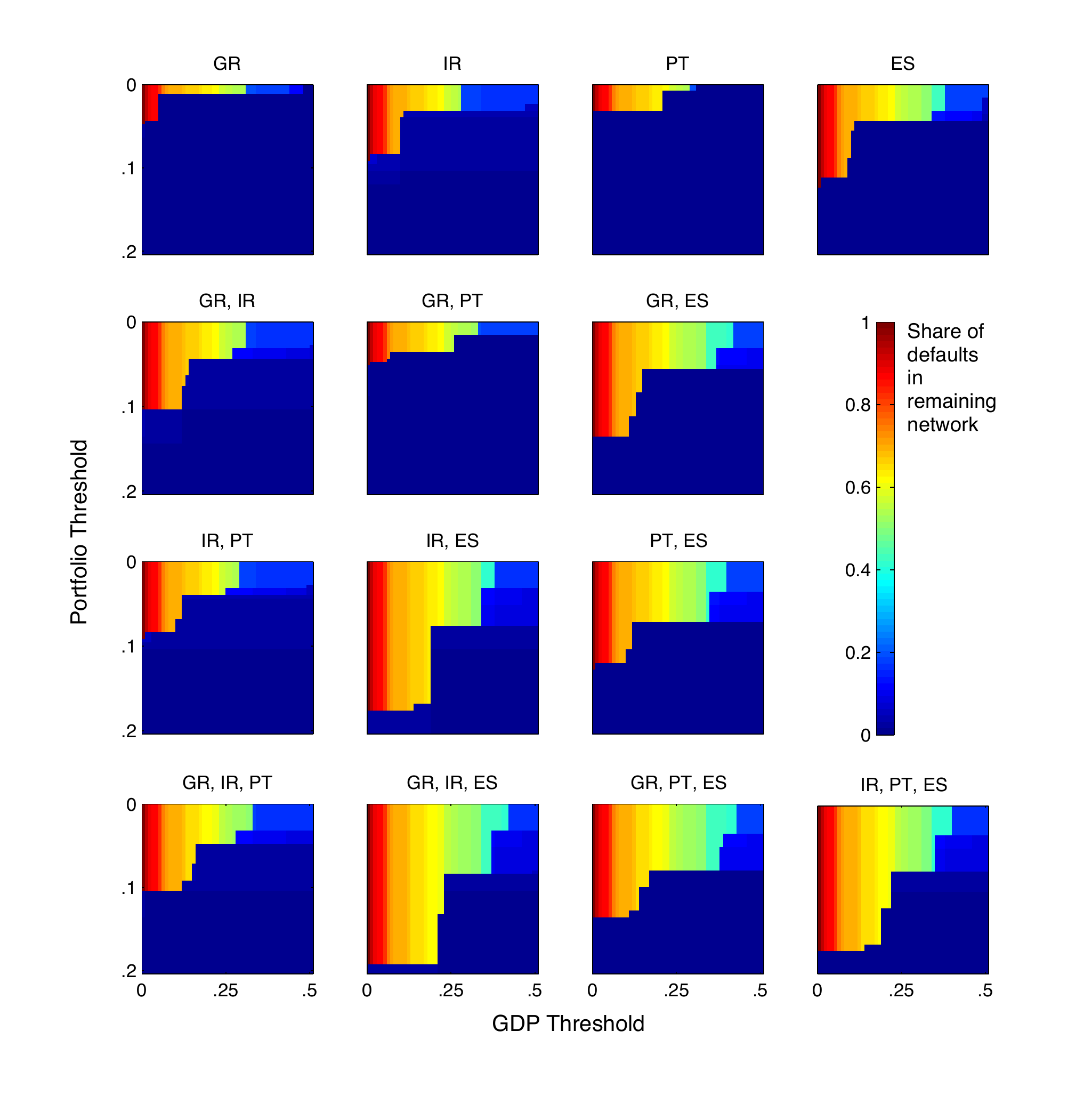}
    \caption{\textbf{Financial Interdependence with respect to PIGS countries.} Plots show the resulting endogenous defaults, given the initial default of any permutation of up to three PIGS countries in 2007 under different threshold specifications:  that is, portfolio threshold $d_1\in [0,.2]$  (with increments of $0.004$) and  GDP threshold $d_2\in [0, .5]$ (with increments of $0.01$). }
    \label{fig:5_Map_for_PIGS_financial_interdependence}
 \end{figure}

%(Pigs Simulations)
Global concerns over the solvency of the PIGS nations motivates us to consider a restricted study of the initial defaults by PIGS countries in 2007. In this we see the different effects of the two thresholds. Figure \ref{fig:5_Map_for_PIGS_financial_interdependence} shows the degree of the network's financial interdependence with respect to any combination of up to three PIGS countries, with the portfolio threshold $d_{1}\in[0,0.2]$ (with increments of $0.004$) and the GDP threshold $d_{2}\in[0,0.5]$ (with increments of $0.01$). For fixed $d_{2}$, the portfolio threshold $d_{1}$ (on the y-axis) in most cases yields only a unique threshold which determines whether or not a specific initial default leads to further defaults, and when it does, the scenario is generally constant. Conversely, for fixed $d_{1}$, we see a more graduated behavior in terms of number of defaults, steadily decreasing as the threshold increases. 
% (perhaps we could elaborate on the implications of this finding... e.g. general robustness w.r.t. $d_2$ vs. the benefits of diversification that is $d_1$)

 %(Pigs Simulations)
Global concerns over the solvency of the PIGS nations motivates us to consider a restricted study of the initial defaults by PIGS countries in 2007.  In this we see the different effects of the two thresholds. Figure \ref{fig:5_Map_for_PIGS_financial_interdependence} shows the degree of the network's financial interdependence with respect to any combination of up to three PIGS countries, with the portfolio threshold $d_1 \in [0, 0.2]$ (with increments of $0.004$) and the GDP threshold $d_2 \in [0, 0.5]$ (with increments of $0.01$). For fixed $d_2$, the portfolio threshold $d_1$ (on the y-axis) in most cases yields only a unique threshold which determines whether or not a specific initial default leads to further defaults, and when it does, the scenario is generally constant. Conversely, for fixed $d_1$, we see a more graduated behavior in terms of number of defaults, steadily decreasing as the threshold increases. 
 % (perhaps we could elaborate on the implications of this finding... e.g. general robustness w.r.t. $d_2$ vs. the benefits of diversification that is $d_1$)
 
 %Figure 4: Heat map for pigs simulations
 
 % Also note decrease of financial interdependence in 2008p. At the end of 2008, financial crisis decreased assets in the system greatly > in LGD model more stability w.r.t.\ d_2, however this does not account for perhaps already occured losses. Suggests one area for improvement. plus ...

 %\emph{(Something along the lines of: Nonetheless, our two approaches to characterize the international financial asset network provide an insightful analytic foundation.)}
\section*{Conclusion}

 We believe that the robustness studies undertaken here are an important first step in the development of metrics for the study of systemic risk in the global economy. The application of error and attack analysis on the CPIS network and its effect on the average shortest path length produce robustness results similar to those of a  scale-free network, indicating robust-yet-fragile structure. Loss-given-default dynamics produce simulations that show an increase from 2001 to 2007 in  network fragility with respect to failure of key countries. The different analytical tools all highlight the key importance of the United States and the centrality of european countries. In general, most simulations support the idea that a failure of the US in 2008 would have had far reaching consequences for the entire network. Similarly, the concerns over the default of Greece seem real as simulations indicate that with the failure of Greece (or any of the PIGS nations) the global economy was one default away from a contagion cascade. Models that assume low thresholds for contagion also predict that the default of a combination of PIGS countries may be similarly severe. The only countries relatively unaffected by such global finical crisis seem to be middle income countries, whose external financial assets are relatively small as a share of their GDP. We believe that these and analogous knock-out studies may be of use in further refining our understanding of the global financial network. Further, more targeted simulations may help informing important policy decisions.

 \bibliographystyle{naturemag}
 \bibliography{myrefs}

\end{document}

%% file: Top_countries.tex
\begin{tabular}{lr|llr|lllr}
\hline\hline
1 Attack&Instances&2 Attacks&&Instances&3 Attacks&&&Instances\\\hline
United States&540&United Kingdom&United States&301&Germany&United Kingdom&United States&200\\
United Kingdom&166&Brazil&United States&280&Brazil&United Kingdom&United States&178\\
Cayman Islands&159&Germany&United States&198&Brazil&Germany&United States&163\\
Germany&142&Turkey&United States&177&Brazil&Turkey&United States&156\\
Luxembourg&142&Indonesia&United States&174&France&United Kingdom&United States&154\\
France&130&France&United States&167&Brazil&Russia&United States&153\\
Brazil&123&Russia&United States&156&Brazil&Poland&United States&148\\
Italy&118&Italy&United States&155&Brazil&Indonesia&United States&147\\
Netherlands&118&South Korea&United States&153&Brazil&Colombia&United States&147\\
Japan&115&Australia&United States&153&Germany&Italy&United States&146\\
\hline\hline
\end{tabular}